\documentclass{article}

\usepackage{PRIMEarxiv}

\usepackage[utf8]{inputenc} 
\usepackage[T1]{fontenc}    
\usepackage{hyperref}       
\usepackage{url}            
\usepackage{booktabs}       
\usepackage{amsfonts}       
\usepackage{nicefrac}       
\usepackage{microtype}      
\usepackage{lipsum}
\usepackage{fancyhdr}       
\usepackage{graphicx}       
\graphicspath{{media/}}     

\title{A Real-time Endoscopic Image Denoising System
}

\author{
  Yu Xing\textsuperscript{1}\footnotemark[1]~~~~Shishi Huang\textsuperscript{1}~~~~Meng Lv\textsuperscript{1}~~~~Guo Chen\textsuperscript{1}~~~~Huailiang Wang\textsuperscript{1}~~~~Lingzhi Sui\textsuperscript{1} \\
  \\
  \textsuperscript{1} Beijing Yi Li Technology \\
  \footnotemark[1]~~\texttt{xingy16@hotmail.com}
}

\begin{document}
\maketitle

\begin{abstract}
Endoscopes featuring a miniaturized design have significantly enhanced operational flexibility, portability, and diagnostic capability while substantially reducing the invasiveness of medical procedures. Recently, single-use endoscopes equipped with an ultra-compact analogue image sensor measuring less than 1mm$\times$1mm bring revolutionary advancements to medical diagnosis. They reduce the structural redundancy and large capital expenditures associated with reusable devices, eliminate the risk of patient infections caused by inadequate disinfection, and alleviate patient suffering. However, the limited photosensitive area results in reduced photon capture per pixel, requiring higher photon sensitivity settings to maintain adequate brightness.  In high-contrast medical imaging scenarios, the small-sized sensor exhibits a constrained dynamic range, making it difficult to simultaneously capture details in both highlights and shadows, and additional localized digital gain is required to compensate. Moreover, the simplified circuit design and analog signal transmission introduce additional noise sources. These factors collectively contribute to significant noise issues in processed endoscopic images. In this work, we developed a comprehensive noise model for analog image sensors in medical endoscopes, addressing three primary noise types: fixed-pattern noise, periodic banding noise, and mixed Poisson-Gaussian noise. Building on this analysis, we propose a hybrid denoising system that synergistically combines traditional image processing algorithms with advanced learning-based techniques for captured raw frames from sensors. Experiments demonstrate that our approach effectively reduces image noise without fine detail loss or color distortion, while achieving real-time performance on Field Programmable Gate Array (FPGA) platforms and an average Peak Signal-Noise-Ratio (PSNR) improvement from 21.16 to 33.05 on our test dataset.
\end{abstract}


\section{Introduction}
\begin{figure}
\includegraphics[width=\textwidth]{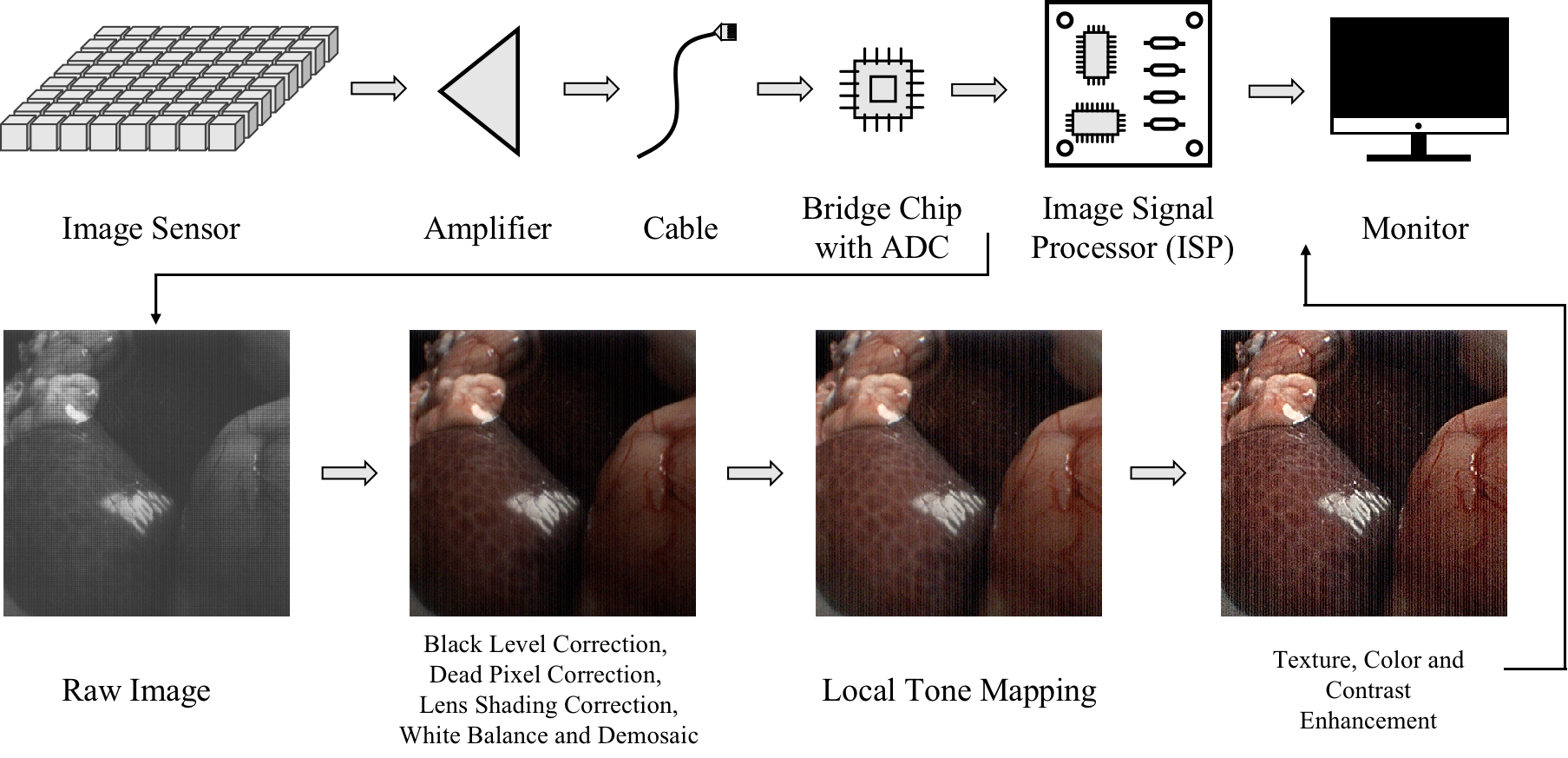}
\caption{(a). Medical endoscope system equipped with an ultra-compact analog image sensor. (b). The overview of an Image Signal Processor (ISP) pipeline.} \label{isp_pipeline}
\end{figure}

The miniaturized design of endoscopes facilitates direct visualization of inaccessible remote tissues through narrow anatomical lumens, bringing revolutionary changes to both clinical diagnostics and biological research. The image sensor is sufficiently small to be embedded at the distal end of the endoscope, eliminating the need for traditional fiber-optic image transmission~\cite{fiber}. For example, SpyGlass DS II System~\cite{spyglass} with 3.5mm distal tip width delivers high-resolution imaging of the biliary and pancreatic duct systems during an endoscopic retrograde cholangiopancreatography (ERCP) procedure, providing critical insights~\cite{spyglass_application} into biliopancreatic lesions (such as stricture, cyst, tumor) and calculi. The simplified hardware design and reduced cost also enable the rapid development of single-use endoscopes~\cite{ambu,pentax}, significantly mitigating the risk of infection associated with inadequate sterilization of reusable endoscopes~\cite{infection}.

However, image noise remains a significant challenge in such endoscope systems. As shown in Fig~\ref{isp_pipeline}, in an endoscopic system equipped with an ultra-compact analog image sensor such as OV6946~\cite{ov6946} or OH0FA10~\cite{fa10}, with package dimensions of 0.95mm$\times$0.95mm, the image sensor initially transforms the optical signal into an electrical signal. This analog signal is then transmitted through an amplifier and a cable to a bridge chip. Within the bridge chip, the analog signal is quantized by an Analog-to-Digital Converter (ADC). Throughout this process, several noise~\cite{noise-model} including shot noise, readout noise, fixed-pattern noise (FPN) and quantization noise are generated. Subsequently, the digital signal, serving as a raw image, is fed into an Image Signal Processor (ISP) for digital image processing. A typical ISP~\cite{isp} incorporates various modules such as auto exposure (AE), black-level correction (BLC), dead-pixel correction (DPC), lens-shading correction (LSC), white balance, demosaicing, local tone mapping (LTM), and texture, color and contrast enhancement, each amplifying the noise introduced by the sensor's physical mechanisms. 

Traditional image processing algorithms~\cite{vbm4d,hdrplus} have demonstrated remarkable effectiveness in processing images corrupted by additive white Gaussian noise or Poisson-Gaussian noise~\cite{mri}. However, the performance significantly deteriorates in extremely dark environments or scenarios characterized by low signal-to-noise ratio~\cite{extreme,see}. With the development of deep learning, algorithms based on Convolutional Neural Network~\cite{pmrid} or Transformer~\cite{swin} have emerged as predominant solutions. Some approaches rely on extensive paired datasets for training~\cite{see}, but the acquisition of data proves particularly challenging in the medical domain. Alternative self-supervised methods~\cite{noise2noise}, which require only noisy data, are incapable of handling spatially structured noise. Currently, the approaches which involve precise modeling of sensor noise and processing in the raw-domain~\cite{raw-denoise} or dual-raw-rgb-domain~\cite{dual-dn} are considered to be practical and functional in image denoising. Processing in the Raw domain avoids the significant gain and nonlinear distortions introduced by ISP modules. Synthetic paired data generated by noise model has been proved to be effective for training~\cite{extreme,raw-denoise,pmrid}. Nevertheless, there remains a scarcity of research focusing on noise modeling and denoising techniques specifically for medical endoscopy systems, which additionally impose stringent real-time processing requirements. To address these challenges, the contributions of this study are as follows:

\begin{enumerate}

    \item A physically-based sensor noise model for medical endoscopy systems, encompassing fixed-pattern noise, periodic banding artifacts, Poisson-Gaussian noise, and quantization noise.
    \item Proposal of a denoising system that integrates a heuristic method for fixed-pattern noise and striping artifact removal and a learning-based approach for residual noise reduction. This system eliminates the need for paired real data and ensure real-time operational capability.

\end{enumerate}

\section{Methods}

In previous studies, the intrinsic noise of sensors has typically been modeled as a combination of Poisson-distributed noise, resulting from the quantum fluctuations of light, and Gaussian-distributed readout noise. However, in practical medical sensors, factors such as circuit design and impedance/phase mismatches during data transmission, particularly in long-distance analog signal transmission, as well as power supply stability and sensor temperature, significantly influence the noise distribution of the output signal. These factors lead to raw images containing not only the typical Poisson-Gaussian noise but also fixed-pattern noise (FPN) and severe periodic banding noise (PBN). In this work, we raise a physics-based noise model and subsequently propose a real-time two-step algorithm to effectively eliminate image noise in raw-domain.

\subsection{Physics-inspired Noise Modeling}

Our goal is to recover a clean image $I$ from corresponding noisy image $I'$. The physics-inspired noise model consists of the following components:
\begin{equation}
\label{summary}
I' = I + N_{shot}+N_{gaussian}+N_{q}+N_{FPN}+N_{periodic}
\end{equation}
As shown in Eq.~\ref{summary} and Eq.~\ref{shot}, shot noise $N_{shot}$ in sensors is the random fluctuation in photon arrival due to quantum effects, following a Poisson distribution where the noise variance is proportional to the signal intensity $I$. The dark current shot noise, sense node reset noise and source follower noise are all considered to be Gaussian-distributed~\cite{noise-model} as $N_{gaussian}$. The Analog-to-Digital Converter transforms continuous analog signals into discrete digital signals, introducing quantization noise $N_q$ that typically follows a uniform distribution $\mu(\lambda)$.
\begin{equation}
\label{shot}
N_{shot}+N_{gaussian}+N_{q} \sim Poisson(I)+\mathcal{N}(0,~\sigma^2)+\mu(\lambda)
\end{equation}
Fixed-Pattern Noise (FPN) consists of dark current FPN and offset FPN. Dark current FPN primarily arises from variations in photodetectors and surface defects caused by semiconductor manufacturing processes~\cite{noise-model}. Consequently, the average dark signal across different pixels is inconsistent. Offset FPN occurs because pixels within the same column or Bayer phase share the same amplifier, the offsets across different amplifiers vary, resulting in column-to-column Offset FPN. Consequently, the FPN of the sensor usually has a spatially-random but fixed-pattern structure. Previous research~\cite{iso-depend} for FPN suggests that FPN is related to the camera's ISO level, with the position and intensity of the noise being stable across different individual sensors of the identical model. However, we have found that FPN is also significantly influenced by exposure time and temperature, and exhibits considerable variation among different individual sensors of the same model. As shown in Eq.~\ref{fpn}, the variables $x$ and $y$ denote the horizontal and vertical coordinates in the image respectively. The FPN at each pixel $N_{FPN}(x, y)$ is characterized by fitting a slope $K(x,y,temp.)$ and an offset $B(x,y,temp.)$, which are influenced by the temperature. The FPN increases proportionally as the analog gain and exposure time $t$ are elevated.
\begin{equation}
N_{FPN}(x,~y) \sim K_(x,~y,~temp.) \cdot analog\_gain \cdot t + B(x,~y,~temp.)
\label{fpn}
\end{equation}
The Periodic Banding Noise (PBN) is caused by impedance/phase mismatches during long-distance transmission of analog signals~\cite{stripe}. Long-distance transmission is often not required by sensors used in mobile and automotive applications, so few researchers have addressed this issue. In Eq. \ref{periodic}, we model the periodic vertical banding noise as a square wave signal superimposed on the original image, where the amplitude is denoted by $\kappa$ and the frequency is represented by $\theta_{x}$. Additionally, the $\kappa$ exhibits variations in response to voltage instability and transmission distance of analog signals.
\begin{equation}
N_{periodic}(x) = \kappa \cdot f(sin(\theta_{x})), \quad f(p)=1~if~p>0~else~-1
\label{periodic}
\end{equation}

\subsection{Denoising System}

Each individual sensor has a unique structure of FPN, even learning-based methods are unable to discern and subsequently eliminate this pattern from a single frame. The FPN here is characterized as additive noise, independent of the intensity of the original signal. Consequently, we can capture dark frames and determine the slope~$K$ and offset~$B$ in Eq.~\ref{fpn} through offline calibration. However, prior to this step, it is imperative to remove the periodic banding noise that also exists in the dark frames.

\subsubsection{Periodic Banding Noise Removal}

\begin{figure}
\includegraphics[width=\textwidth]{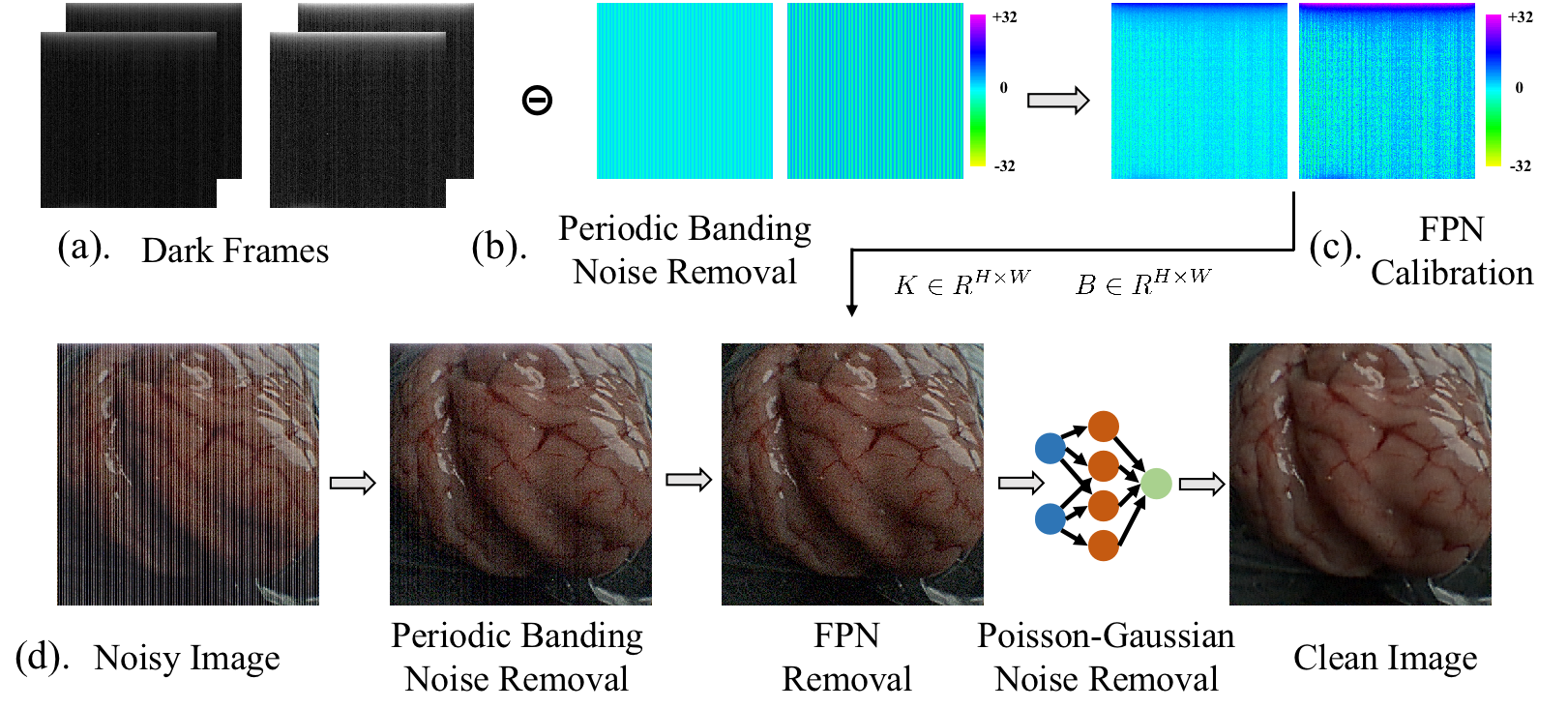}
\caption{(a). Capture several sets of sequential dark-field frames with different analog gains and exposure time. (b). Remove periodic banding noise in dark frames according to Eq.~\ref{kappa}. (c). Calibrate the slope and bias of FPN with respect to the analog gain and exposure time. (d). The overview of our denoising system. PBN and FPN will be removed firstly, and the image with Poisson-Gaussian Noise is fed into a learning-based Convolutional Neural Network. Algorithms above are implemented in raw-domain, images shown here are demosaiced for better visualization.} \label{denoise_pipeline}
\end{figure}
In the sensor utilized~\cite{ov6946} in this work, the square wave depicted in Eq.~\ref{periodic} representing Periodic Banding Noise exhibits a fixed period of 4 pixels along the horizontal axis of the image. Subsequently, we have to determine the amplitude and initial phase of the square wave. For each row $y$ in the image, we employ the method in Eq.~\ref{kappa} to determine the amplitude $\kappa$. Within a row in the image, pixels that are adjacent at intervals of 2 share the same Bayer phase, with each being overlaid by either the positive or negative segment of the square wave. In the flat regions, the amplitude of the square wave can be obtained by taking the absolute value of the difference of $I'(i+2,~y)$ and $I'(i,~y)$, and we can deduce the initial phase of the square wave by examining the sign of $I'(i+2,~y) - I'(i,~y)$ at different positions. The weight $w$ is used to determine whether the region is flat, which can be simply obtained using a thresholding method~\cite{rough} in Eq.~\ref{weights}. Ultimately, we derive the final result $\kappa$ by averaging the amplitude $\kappa_y$ across each row.
\begin{equation}
4 * \kappa_y = \sum_{i} (|I'(i-2,~y) + I'(i+2,~y) - 2 \cdot I'(i,~y)| \cdot w_{i,y})~/~\sum_{i}w_{i,y}
\label{kappa}
\end{equation}
\begin{equation}
w_{i,y} = 1~if~|I'(i+2,~y)-I'(i-2,~y)| < \theta~else~0
\label{weights}
\end{equation}

\subsubsection{FPN Calibration and Removal}

The temperature has weak influence on the sensors utilized in this study, we temporarily disregarded the impact of temperature on FPN. We captured several sets of consecutive dark frames, with the analogue gain and exposure time set to $analog\_gain_i$ and $t_i$, respectively. After subtracting the PBN, we computed the temporal average image $\bar{I'}_i$ for each set. The temporal averaging process effectively removes shot noise and gaussian noise from the dark frames, leaving behind only the residual FPN. Based on Eq.~\ref{fpn}, we can calibrate the slope~$K$ and offset~$B$ of the FPN at each pixel~$(x, y)$ using $\bar{I'}_i$. In the denoising process, once the current frame's analogue gain and exposure time are determined, the FPN for the current frame can be directly calculated.

\subsubsection{Poisson-Gaussian Noise Removal}

After the elimination of Periodic Banding Noise and FPN, the challenge transitions into Poisson-Gaussian noise reduction, which has been well explored. We captured a large number of grayscale images~\cite{cmu} to model the distribution of Poisson and Gaussian noise under different analogue gains. Leveraging the calibrated noise models, we generated substantial synthetic paired images and adopted a learning-based approach utilizing a U-Net~\cite{unet} architecture. To enhance computational efficiency, we adopt a strategy inspired by PMRID~\cite{pmrid}, replacing convolutional operations in the encoder and decoder with a combination of depth-wise and point-wise convolutions.

\section{Experiments and Results}

We validated our denoising system on two OV6946~\cite{ov6946} sensors. The OV6946 exhibited significant PBN and notable FPN. We captured 200 sets of paired data using the OV6946, which served as the test datasets for evaluation.

\subsection{Test Dataset Construction}
We observed that when the analog signal transmission distance, analog gain amplitude, and exposure time were fixed, along with the use of a more stable power supply and no adjustments to the LED light of the endoscope system during imaging, the sensor exhibited stable FPN and PBN. Therefore, under these fixed conditions, we first captured 128 dark-field images $d_i$ and averaged them to obtain the FPN and PBN under the current conditions, denoted as $\bar{d}$. Next, we captured 128 images $l_i$ under normal illumination and averaged them to obtain $\bar{l}$, thereby mitigating Gaussian and Poisson noise. The noise-free image under the current conditions was then derived as $\bar{l}-\bar{d}$, completing the creation of one pair of data. Subsequently, we varied the exposure time, analog gain, and imaging targets to ensure that the test set covered a wide range of signal-to-noise ratio (SNR) scenarios.

\subsection{Training Dataset Construction}
As shown in Fig.~\ref{denoise_pipeline} (d), after the PDN and FPN are removed, we have transformed the problem into a Poisson-Gaussian noise removal by a learning-based model. Learning-based methods require a substantial amount of training dataset, but obtaining real-world medical data is challenging. To address this, we explored the use of synthetic data for training and will subsequently validate the effectiveness of this approach. Firstly, we utilized $l_i-\bar{d}$ to calibrate the Poisson-Gaussian noise strength of the sensor based on methods described in the literature~\cite{cmu}. We then applied noise of the corresponding intensity to the ground truth provided by the SID dataset proposed in~\cite{see}, generating synthetic data to serve as the training set. We selected 200 high-quality long-exposure RAW images from the SID dataset that exhibited no significant noise. During training, we employed basic data augmentation techniques, such as affine transformations and contrast enhancement, to improve the robustness of the model.

\begin{table}
\centering
\caption{PSNR evaluation of our denoising system, and qualitative comparison with other approaches.}\label{tab1}
\begin{tabular}{|c|c|c|c|c|c|c|c|c|}
\hline
Sensor & \multicolumn{8}{|c|}{OV6946}\\
\hline
Analog Gain & \multicolumn{2}{|c|}{Low}  & \multicolumn{2}{|c|}{Medium} & \multicolumn{2}{|c|}{Large} & \multicolumn{2}{|c|}{All} \\
\hline
Metric & PSNR & SSIM & PSNR & SSIM & PSNR & SSIM & PSNR & SSIM \\
\hline
Noisy&~22.13~ &~0.44~ &~22.29~ &0.43~ &~19.81~ &~0.33~ &~21.16~ &~0.39~ \\
w/o PBN&30.89&0.79&29.22&0.73&27.44&0.63&28.92&0.70\\
w/o PBN, FPN &31.83&0.80&30.05&0.73&27.89&0.63&29.62&0.71\\
Ours(Synthetic)&33.86&0.90&33.45&0.88&32.27&0.84&33.05&0.87\\
Ours(Overfit)&34.45&0.91&33.84&0.89&32.46&0.84&33.41&0.87\\
\hline
CycleISP~\cite{cycleisp}&33.21&0.91&33.39&0.90&32.67&0.86&33.02&0.89\\
SwinIR~\cite{swinir}&33.95&0.89&33.88&0.89&32.86&0.85&33.45&0.87\\
\hline
\end{tabular}
\end{table}

\subsection{Denoising System Performance Validation}
To validate the effectiveness of our noise model and denoising system, we quantified the impact of each individual component in the pipeline by peak signal-noise-ratio (PSNR) and structural similarity (SSIM).
\subsubsection{Ablation Studies}
\begin{figure}
\includegraphics[width=\textwidth]{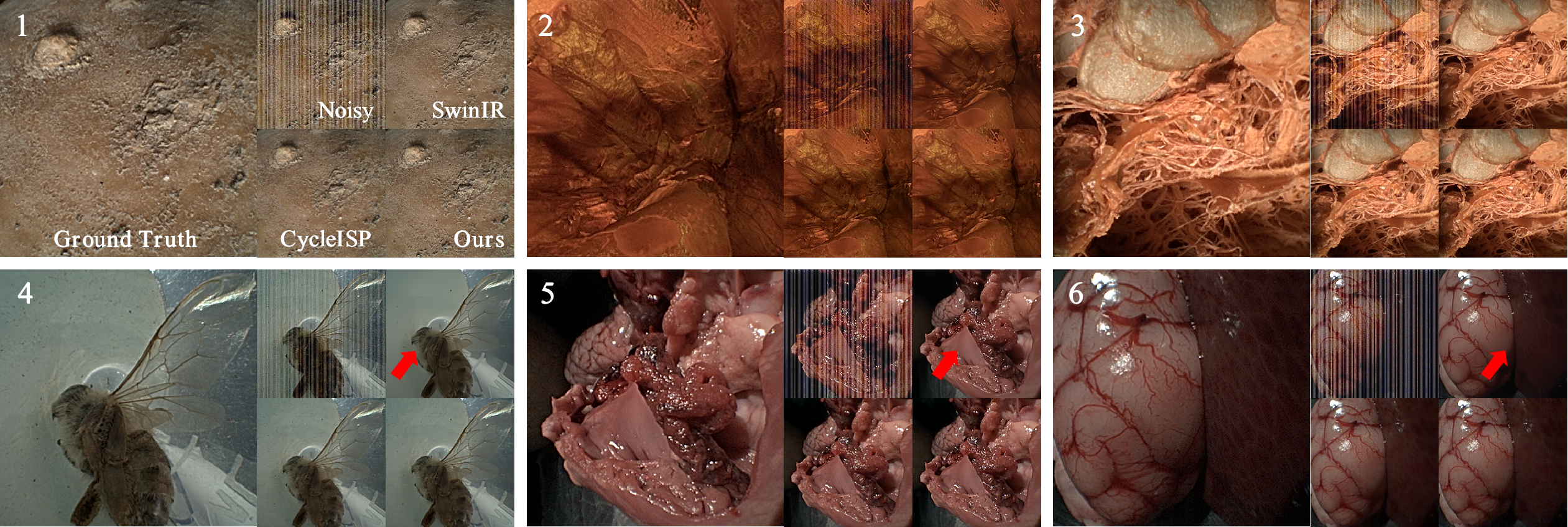}
\caption{Visual results on our real test dataset.} \label{demo}
\end{figure}
As shown in the table.~\ref{tab1}, we evaluated the performance of the denoising system under various analog gain settings and signal-to-noise ratio (SNR) conditions. For example, in low-SNR scenarios with high analog gain, the PSNR increased from an initial value of 19.81 to 27.89 after PBN and FPN removal. Subsequently, with the application of the learning-based denoising model, the PSNR was further improved to 32.27. It is noteworthy that we also directly trained the model on the real test dataset during the experiments, resulting in an overfitted model, which unsurprisingly achieved the best performance. The performance gap between the model trained on synthetic data and the overfitted model was minimal, further validating the accuracy of our modeling approach.
\subsubsection{Discussion}
We provide several visual results on our real test dataset in Fig.~\ref{demo}. We can observe that the ground truth obtained using our proposed method exhibits almost no noticeable noise. Additionally, as shown in the Table.~\ref{tab1}, we compared our results with pretrained CycleISP~\cite{cycleisp} and SwinIR~\cite{swinir}. CycleISP, which leverages a noise model to guide the denoising process, achieved results very close to ours when using the noise model calibrated in this study. SwinIR demonstrated superior PSNR and produced visually impressive results on indoor object images. However, in medical-like scenarios, it introduced noticeable artifacts indicated by the red arrows in the Fig.~\ref{demo}, such as cartoon-like effects, color shifts, and excessive smoothing of fine structures like blood vessels and tissues. Here, we do not intend to assert which method is superior. The primary reason for the performance gap lies in the discrepancy between the noise distribution in the training data and that in real-world images. Particularly in the medical field, it is crucial to meticulously handle noise models and training data, otherwise color deviations, loss of details, and artificial textures can be unacceptable in medical applications. 

Additionally, to improve computational ability and increase memory access efficiency, we have quantized~\cite{compression} all 32-bit floating-point intermediate results and parameters into 12-bit fixed-point numbers without performance drop, and engineered a systolic array~\cite{dnnvm} on an FPGA platform to implement the operations, capable of achieving a performance of 30 frames per second.

\section{Conclusion}

In this paper, we propose a comprehensive physics-based noise model for medical endoscope systems. Building upon this noise model, we introduce a denoising system for RAW images, capable of addressing various types of noise independently without requiring real medical data. Furthermore, leveraging this noise model, we constructed a test dataset to evaluate denoising performance across different SNRs, achieving an average PSNR improvement from 21.16 to 33.05. This denoising system can be naturally integrated into current image signal processing (ISP) pipeline of medical endoscopes, laying a solid foundation for more advanced post-processing and high-level applications.

\bibliographystyle{unsrt}  

\end{document}